# Lying mirror using structured surfaces

Yuhang Li[1,2,§], Shiqi Chen[1,2,§], Bijie Bai[1,2], and Aydogan Ozcan[*,1,2,3]

[1]Electrical and Computer Engineering Department, University of California, Los Angeles, CA, 90095, USA.

[2]California NanoSystems Institute (CNSI), University of California, Los Angeles, CA, USA.

[3]Bioengineering Department, University of California, Los Angeles, 90095, USA.

[§] Equal contribution

[*]Correspondence: Aydogan Ozcan. Email: ozcan@ucla.edu

## Abstract

We introduce an all-optical system, termed the "lying mirror", to hide input information by transforming it into misleading, ordinary-looking patterns that effectively camouflage the underlying image data and deceive the observers. This misleading transformation is achieved through passive light-matter interactions of the incident light with an optimized structured diffractive surface, enabling the optical concealment of any form of secret input data without any digital computing. These lying mirror designs were shown to camouflage different types of input image data, exhibiting robustness against a range of adversarial manipulations, including random image noise as well as unknown, random rotations, shifts, and scaling of the object features. The feasibility of the lying mirror concept was also validated experimentally using a structured micro-mirror array along with multi-wavelength illumination at 480, 550 and 600 $nm$, covering the blue, green and red image channels. Furthermore, we created a broadband lying mirror that operates across a continuous spectral range, enhancing its adaptability under diverse illumination conditions. This framework showcases the power of structured diffractive surfaces for visual information processing and might find various applications in defense, security and entertainment.

## Introduction

Lying mirrors, often referred to as deceptive or distorted mirrors, are an intriguing subject within optical sciences and engineering, leveraging subtle curvatures of mirrors to manipulate visual perception. This phenomenon is grounded in the manipulation of light reflection, wherein variations in mirror curvature alter the perceived dimensions of reflected object waves. Unlike standard planar mirrors, which reflect light uniformly and maintain the object's true proportions, lying mirrors use deliberate concave or convex shapes to distort the reflected wave that carries object information. The optical behavior of these mirrors is dictated by the curvature's impact on light rays: concave mirrors cause light to converge, resulting in a reflection that elongates and slims the object, while convex mirrors cause divergence, producing a broader, shorter appearance. These lying mirrors can be regarded as the apertures of an imaging system, which can obscure an object's perception, misleading the observers. This subtle yet effective alteration of reflected images has various applications in entertainment and experimental psychology, where they can be used for studying perceptual distortions[1–4]. Furthermore, lying mirrors, which manipulate visual perception through optical distortions, may also present applications in defense and security, including camouflage and deception[5–9]. For example, these mirrors could contribute to enhancing camouflage by altering the appearance of equipment, vehicles, or personnel under specific viewing conditions, making them appear differently shaped, smaller, or larger, thus confusing adversaries and delaying target identification[10–14]. In structured optical environments such as viewing ports, surveillance windows, or underwater detection systems, the technology may also help disrupt adversarial imaging techniques, reducing the effectiveness of target identification. Additionally, lying mirrors can be used to create disorienting environments that confuse adversaries by altering the perception of space[15–18]. Such a lying mirror framework could also be valuable in anti-observation tactics, creating visual ambiguity that complicates accurate targeting by opposing forces[19–21]. In scenarios where coherent illumination is used for reconnaissance, lying mirrors might be integrated into semi-transparent materials or optical interfaces to alter the perceived appearance of objects, potentially making them seem differently shaped or visually obscured. By manipulating reflections to obscure true positions, lying mirrors can provide an additional layer of protection for personnel and equipment[22]. Such strategic uses of lying mirrors in defense and security showcase how optical engineering can be leveraged to manipulate perception, enhance concealment, and improve situational awareness in critical operations.

Here, we present an all-optical lying mirror concept that leverages programmed light diffraction through a passive structured surface to optically transform arbitrary input messages and objects into visually misleading output patterns, as shown in Fig. 1a. Our lying mirror design consists of a standard reflective mirror and a trainable diffractive layer, which was optimized using deep learning, spanning only $53.3\lambda$ axially, where the $\lambda$ is the illumination wavelength. This transformation process, from input to output images, is purely performed in the optical domain through the propagation and programmed diffraction of light. As a result, no electronic or digital processing is required, and once the diffractive layer is fabricated, the system operates without any external power consumption. We demonstrated the effectiveness of this lying mirror design through various models trained on different image datasets. Numerical blind testing results validated the feasibility of our approach, with the diffractive lying mirror consistently converting arbitrarily selected unseen, new input images into uniform and ordinary-looking "dummy" images at the output field of view (FOV). Importantly, these lying mirrors were designed to work for infinitely many different input objects, as opposed to working for only a few input structures/objects. Therefore, external generalization to different types of input distributions and randomly selected objects was rigorously demonstrated to validate the success of our lying mirror designs. The system's resilience was also evaluated under various conditions, including random rotations, shifts and scaling of the input objects, as well as varying levels of input noise, confirming the robust performance of lying mirrors against such unknown and random perturbations. Furthermore, we experimentally demonstrated the proof of concept of the lying mirror framework using a structured micro-mirror array under multi-wavelength visible illumination at 480, 550 and 600 $nm$, covering the blue, green and red image channels. We believe that lying mirrors based on diffractive visual information processing might have various applications in defense, security and entertainment.

## Results

### Design of a lying mirror to conceal optical information

Initially, we numerically demonstrated the framework of lying mirrors for optical information concealment using three different datasets: Fashion-MNIST[22], MNIST[24], and QuickDraw[25]. As depicted in Fig. 1a, our lying mirror configuration included a diffractive layer paired with a reflective standard mirror, designed to transform infinitely many input images into deceptive target patterns, i.e., "dummy" images at the output. The lying mirror was engineered to optically conceal input data entirely, functioning under spatially coherent illumination at a wavelength of $\lambda$. The diffractive layer consisted of 120×120 phase-modulating features, each ~$\lambda$/2 in lateral size (detailed in the **Supplementary Information**). Illustrated in Fig. 1c, this setup processes an arbitrarily selected input image through the lying mirror, producing an output image that visually deceives observers by appearing as a different object than what was actually presented. This desired optical concealment is supposed to work for infinitely many different input objects, as opposed to working for only a few input structures; thus, generalization to different types of input distributions is a requirement for the performance of a lying mirror. Stated differently, the concepts outlined here should not be confused with the conversion of a few different and known input objects/structures into a different output image since the lying mirror designs reported in this work do not memorize input structures and would work for any randomly selected input information, as detailed below.

In our numerical demonstrations, as shown in Fig. 2, we trained each lying mirror with a specific image dataset to consistently generate a "dummy" output message, intended to be reproduced by the lying mirror for all input variations. For example, Model 1 of Fig. 2 was designed to optically convert a series of amplitude images of various items, including different "shirts", "backpacks" and "shoes" into a consistent output of “bags”. We employed a deep learning-based optimization approach, which involved feeding the input image dataset through the lying mirror, calculating the corresponding outputs, and comparing them to the predefined target output images (see the **Methods** section and the **Supplementary Information**). For all these three lying mirror designs, during this iterative training and optimization process, the diffractive layers’ phase modulation values were optimized using a Pearson Correlation Coefficient (PCC)-based loss function (detailed in the **Methods** section), which aimed at maximizing the structural similarity between the lying mirror’s output and the predefined target misleading images, i.e., the dummy image for each lying mirror design.

After the training, we performed a numerical evaluation of each lying mirror design using randomly selected images from the corresponding test dataset, which were never used during the training phase. As shown in Fig. 2, the lying mirror adeptly altered various input images—clothing, handwritten digits, or animals—into uniform, misleading patterns (tailored to the specific target of the dummy image), effectively concealing the original content. The PCC values between the output and target/desired images were 0.97, 0.97 and 0.95 for these three models designed using Fashion-MNIST, MNIST, and QuickDraw image datasets, respectively, demonstrating the effectiveness of our lying mirrors in obscuring input information.

In addition to these internal generalization tests, where the lying mirrors were tested with randomly selected new/unseen objects from the same image dataset used during training, we also assessed the external generalization capability of the same lying mirrors using image datasets different from the training data distribution. For example, Model 1, initially trained with the Fashion-MNIST dataset, was blindly tested using the MNIST, QuickDraw and ImageNet datasets. These successful external generalization tests, visualized in Fig. 3, indicate that our lying mirrors can consistently camouflage various randomly selected input images into the designated target images, despite the fact that these inputs are significantly different from the training image data and never encountered during training or design phase of the lying mirror. The average PCC values for Model 1 tested on MNIST, QuickDraw and ImageNet images were 0.84, 0.85 and 0.92, respectively. It is worth noting that the ImageNet dataset images are more diverse and distinct compared to those from the other two datasets, highlighting the lying mirror’s external generalization ability to process inputs that are out-of-distribution compared to the training dataset. Although there was some performance degradation compared to the blind testing results with internal generalization data, the lying mirror still successfully produced deceptive images matching the desired dummy image structure at the output FOV. Similar analyses were also applied to Model 2, trained with MNIST, and Model 3, trained with QuickDraw, revealing the successful external generalization of these lying mirror designs. Each model effectively converted respective inputs into uniform target images, validating the adaptability and effectiveness of our lying mirror designs. We also trained three additional lying mirror designs that transformed the Fashion-MNIST, MNIST, and QuickDraw datasets into a target output image selected from ImageNet using a larger diffractive layer. These results, shown in Supplementary Fig. S1, demonstrate the capability of our lying mirror framework to handle more intricate image transformations.

**Evaluation of lying mirror performance under various image perturbations and noise conditions**

In addition to showcasing the primary functionality of our lying mirror, we also evaluated the system's resilience against various image distortions. We subjected the Model 1 trained with Fashion-MNIST, to random and unknown image rotations, shifts and scaling (detailed in the **Supplementary Information**). As illustrated in Fig. 4, despite random alterations in the input image orientation, location, and size, the lying mirror consistently converted the input images into the target dummy image 'bag' with high fidelity, resulting in average PCC values of 0.95, 0.90 and 0.94 for rotation, shifting and scaling, respectively, across all tested objects.

Furthermore, another critical aspect of our robustness assessment involved testing the system's performance in the presence of random noise. Specifically, we examined the lying mirror's performance when the Gaussian noise was added to the input images. For this evaluation, random Gaussian noise ($\varepsilon$) was introduced into the input images to be processed by the trained Model 1. The noise followed a normal distribution:

$$\varepsilon \sim \boldsymbol{\mathcal{N}}(\mu = 0, \sigma^2) \qquad (1).$$

After adding random noise, any negative values in the resulting images were clipped to zero. As depicted in Fig. 5, the lying mirror underwent testing at two levels of noise, with random image perturbations increasing from $\sigma = 0.05$ to $\sigma = 0.1$. Despite a noticeable decrease in the output image fidelity as the noise level increased to $\sigma = 0.1$, the lying mirror could still successfully transform the input images into concealed dummy images, fulfilling its design function.

These tests collectively highlight the adaptability and resilience of our lying mirror designs in managing diverse and complex perturbations and noise scenarios. Whether facing random structural image distortions or noise, the system consistently maintained its capability of information concealment into dummy output images, even though the original model was not trained to handle such variations in the input image orientation, location, size, or noise.

**Analysis of lying mirror performance with respect to input image sparsity**

To further investigate other factors influencing the image transformation performance of lying mirrors, we quantified the ratio of the active pixels—defined as the proportion of the pixels within the input FOV that exceed a specific intensity threshold (empirically set as 0.1). In addition to three datasets used earlier, i.e.,

MNIST, Fashion MNIST and QuickDraw, we also evaluated the lying mirror model's performance using binary random input patterns and random patterns with a uniform distribution $\mathbb{U}[0, 1]$ – these constitute external test datasets never used during the training of our lying mirrors. Across all three lying mirror models, we observed that as the ratio of the active pixels within the input FOV increased, the output PCC values also increased, irrespective of the dataset used for training (Fig. 6a). This trend suggests that a higher count of active pixels generally enhances the model's ability to conceal input images. We also employed the Peak Signal-to-Noise Ratio (PSNR) metric to assess the output image quality (see the **Methods** section), with a similar performance observed in Supplementary Fig. S2. Additionally, tests under special illumination conditions—point source and uniform illumination—further confirmed the same trend. Figure 6b visualizes the input and output images for Model 3, with the corresponding PCC and PSNR values also indicated in each case.

Furthermore, all three lying mirror models demonstrated strong internal generalization capabilities, evidenced by the high PCC values when tested with the dataset on which they were trained. Although the PCC values for external generalization were lower than those for the internal datasets, they were still significantly better than those for random noise input patterns, as shown in Fig. 6. This indicates the importance of structured input patterns in the models' ability to perform the functions of a lying mirror. During the training phase, the lying mirror effectively learns the spatial distribution of information within the images to be concealed – although it does not overfit to the specific training dataset as shown in our earlier external generalization test results (see Figs. 1c, 3 and 6). The observed improvement in the performance of a lying mirror with an increase in the active pixel ratio underscores the critical role of information density in effective image concealment; a denser input image FOV effectively utilizes more of the spatially-varying coherent point spread functions that are optimized between the input and output FOVs, which can be considered as the engineering degrees of freedom of our lying mirror design. These insights highlight the necessity for lying mirror models to be trained on diverse image datasets to capture more generalized features in order to broaden their effectiveness across various image types.

**A variant of the lying mirror framework: structured mirror-based design**

The architecture of a lying mirror can be further simplified into a structured mirror, as illustrated in Fig. 1b. In this configuration, an optimized diffractive layer is integrated onto a standard reflective surface, specifically engineered to modulate the phase of the incident light waves. In other words, the axial distance between the diffractive layer and the standard reflective mirror in the earlier design is removed in this alternative lying mirror configuration. Using the training method previously described, we developed three structured lying mirrors designed for the Fashion-MNIST, MNIST, and QuickDraw datasets, respectively. As depicted in Fig. 7, these optimized structured mirrors could transform various input images into their corresponding target "dummy" images – as desired. However, compared to the previously introduced lying mirror designs, a performance drop was observed with these structured lying mirrors, achieving PCC values of 0.87, 0.84, and 0.79 for the Fashion-MNIST, MNIST, and QuickDraw datasets, respectively. This relative decline in the output performance of structured lying mirrors can be attributed to the elimination of the axial gap between the structured diffractive layer and the standard reflective mirror, which resulted in a loss of the system's ability to effectively multiplex complex wave modulation by propagating optical waves in a double-pass configuration through the same optimized diffractive layer. Additional results for various axial distances between the optimized diffractive layer and the standard reflective mirror are provided in Supplementary Fig. S3, which shows that performance increases as the axial distance is increased. Nonetheless, the structured mirror-based lying mirror forms an alternative and simpler design that reduces the stringent alignment requirements, particularly in the axial direction, by the integration of the diffractive layer with the mirror at the fabrication stage; this design choice also enhances the overall robustness of the lying mirror device as this integration simplifies the assembly process. Moreover, by combining two components into a single entity, the structured mirror minimizes optical material losses and improves the stability of the system, potentially making it suitable for applications where space and reliability are critical.

**Experimental demonstration of a lying mirror at the visible part of the spectrum**

We demonstrated an experimental implementation of our lying mirror trained with the MNIST dataset using a structured micro-mirror array-based system operating in the visible band, as shown in Fig. 8. A spatial light modulator (SLM) was used to display the randomly selected phase input objects to be concealed, while a phased light modulator (PLM) composed of an array of MEMS-based micro-mirrors functioned as the structured mirror that is optimized, transforming the input images into the target output images (detailed in the **Methods** section and the **Supplementary Information**). The structured mirror, comprising $150 \times 150$ diffractive features, was optimized to process phase-only input images while synthesizing intensity images at the output. It was optimized to convert randomly selected input images under multi-wavelength illumination at $\lambda = 480,\ 550$ and $600\ nm$ to a predefined "dummy" output, i.e., a handwritten digit "8". After its training, the resulting design was displayed on the PLM, forming the structured lying mirror to modulate the incident light, as depicted in Fig. 8b.

We randomly selected 10 different MNIST images that were not used during the training of the lying mirror for the blind testing, as visualized in Fig. 8c. The illumination light, carrying the object information, was modulated and reflected by the structured mirror and then captured by a CMOS camera. The captured intensity was cropped to the central region and normalized to serve as the output from the lying mirror. Figure 8c summarizes the experimental results, where these three illumination wavelengths ($\lambda = 480,\ 550$ and $600\ nm$) were used either sequentially or simultaneously. The sequentially captured image intensities for the three illumination wavelengths and the snapshot multi-color imaging results show that all the handwritten digits (never used before) were successfully transformed into the target dummy handwritten digit '8', experimentally validating the feasibility of our multi-wavelength (RGB) lying mirror to conceal input information of objects.

Beyond the RGB lying mirror design demonstrated above, we further designed a lying mirror for broadband operation, trained across a continuum spanning $520nm - 570nm$ range (see the **Supplementary Information** for training details). The schematic of the experimental setup is shown in Fig. 9a, where two SLMs were used as the input plane and the lying mirror, respectively. We evaluated the performance of this design across various wavelengths selected from $500nm$ to $600nm$, extending beyond the design/training spectral range. As shown in Fig. 9b, the results indicate that the lying mirror achieved high-quality concealment of input images within its training spectral range. Outside this region, the output images still maintained PCC values above 0.85 over a spectrum of $500 - 600\ nm$, demonstrating generalization to input wavelengths that remain outside of the training spectral range (see Fig. 9b). Additionally, we experimentally tested this design under different wavelength illuminations using 10 randomly selected MNIST images, as shown in Fig. 9c. All the input images were successfully transformed into the target dummy handwritten digit '8', experimentally validating the lying mirror's ability to operate across a continuum spanning $500 - 600\ nm$; see Fig. 9c.

It is worth noting that in these experimental designs, we used a flat dispersion model, i.e., the refractive index was assumed to be constant across all wavelengths. As a result, the phase delays were solely governed by the wavelength-dependent phase kernel, rather than material dispersion-related effects. The successful experimental demonstrations of both multi-wavelength and broadband performance of lying mirrors provide direct evidence that the functionality of the lying mirrors arises from the learned diffractive surfaces enabled by our deep learning-based optimization strategy, rather than from any intrinsic material properties.

## Discussion

This work introduced a lying mirror framework that transforms input images into misleading patterns for information concealment, showcasing the capabilities of all-optical diffractive visual processors as a secure and efficient method for data/image/object camouflage. The system's robustness was validated under various forms of attacks, including random and unknown rotations, shifts, scaling variations, and image noise. Moreover, the lying mirror designs performed very well even with datasets for which it was not originally trained, highlighting its adaptability and potential for widespread applications. We also developed a variant of the lying mirror framework that integrated the diffractive layer with a standard mirror to enhance the system's compactness and simplicity. Additionally, the performance of the structured lying mirror was successfully tested under multi-color visible illumination, confirming the practical feasibility of diffractive lying mirrors.

The lying mirror offers a new approach to optical deception, distinguishing itself from conventional reflective surfaces through its broadband operation, adaptability to various input images, and polarization insensitivity. As demonstrated in our experiments (Figs. 8 and 9), the system functions effectively across multiple wavelengths, making it resilient to varying lighting conditions. Additionally, although trained on specific image datasets, it generalizes well to input images significantly different from the training data, camouflaging them into designated target images (Fig. 3). This adaptability contrasts with conventional techniques that typically require predefined input-output mappings, limiting their flexibility.

The design of the optical diffractive layer involves critical trade-offs between the number of diffractive features and pixel size. Increasing the total number of diffractive features enhances the degrees of freedom available for performing a desired optical transformation and improves the output image fidelity, as shown in Supplementary Fig. S4. As for the pixel size, smaller pixels, approaching $\sim\lambda/2$, offer finer control over the light field and achieve better performance, as shown in Supplementary Figs. S5-S6; however, such smaller diffractive features can also pose challenges in fabrication and optimization. Larger pixels can reduce the capability of the lying mirror to conceal the input information effectively, as shown in Supplementary Fig. S6. Nonetheless, this performance loss due to a large pixel size can be mitigated by increasing the axial distance between the structured diffractive surface and the reflective mirror, as demonstrated in Supplementary Fig. S6. Additionally, our lying mirror exhibits sensitivity to the output plane's axial position, with the output image quality degrading with deviations from the optimal position (see Supplementary Fig. S7). To enhance the robustness of a lying mirror against such misalignments, we introduced a "vaccination" strategy during the training[26], deliberately incorporating image plane position shifts. As demonstrated in Supplementary Fig. S7, this vaccination approach allows the system to maintain its performance with a PCC of > 0.6 over a wide range of axial misalignments, ensuring reliable operation even when the image plane deviates from its ideal position. These findings underscore the importance of balancing design parameters to achieve optimal and robust system performance.

We also conducted an adversarial attack on the lying mirror design by employing a digital neural network trained on a set of paired input-output images (see Supplementary Fig. S8). Our evaluations involved adversarial attacks using neural networks trained on varying sample sizes, which revealed that a limited training set could not recover the finer details of the input objects (Supplementary Fig. S8c). Notably, with 500 training examples (i.e., paired input-output images), finer features began to emerge, and with ~10,000 training samples, the adversarial network produced high-quality reconstructions of the input images, capturing intricate details. Additionally, we investigated a jointly trained system in which both the lying mirror and the digital neural network were optimized simultaneously (Supplementary Fig. S9). This integrated approach enables the system to capture complex data relationships, resulting in more robust and precise reconstructions of the input objects compared to the adversarial attacks demonstrated in Supplementary Fig. S8.

The lying mirror framework consists of an isotropic diffractive layer and a reflective mirror, which are polarization-insensitive. These features make the lying mirror a promising candidate for applications in camouflage, imaging, and optical information processing, offering a robust and adaptable solution for various desired optical transformations.

Although our primary results are based on spatially coherent illumination, the design of lying mirrors can also be adapted for spatially incoherent or partially coherent illumination schemes, allowing for operation under diverse lighting conditions, including natural light or light-emitting diodes[27–29]. By simulating random phase profiles and averaging the intensity over time, we can effectively degenerate a coherent system into an incoherent or partially coherent one. This allows the modeling of the propagation and the modulation of the diffractive layer or the structured mirror surface under spatially incoherent or partially coherent illumination using the angular spectrum approach. Therefore, using a similar training process as outlined in this work, we can design diffractive lying mirrors that function under incoherent or partially coherent conditions. As shown in Supplementary Fig. S10, three lying mirrors were trained under partially coherent light with different phase correlation lengths (see the **Methods** section for details), all of which successfully concealed the input images by transforming them into a predefined ordinary image.

Additionally, we demonstrate that the lying mirror maintains its functionality across multiple viewing angles. To showcase this, we trained a lying mirror using only 0° viewing angle at its output, as we did in earlier designs, and blindly tested its generalization at ±5° viewing angles. As shown in Supplementary Fig. S11, the same deceptive output image was observed across a range of viewing angles (±5°), even when the input images varied. Output images at ±5° viewing angles revealed a similar performance as the 0° viewing angle, quantified by the PSNR and PCC values in each viewing angle; Supplementary Fig. S11 showed that the mean PSNR and PCC values at ±5° viewing angles only decreased by ~8.2% and ~1.6%, respectively, compared to the 0° viewing angle. While these results confirm that the lying mirror preserves its deceptive image transformation capability under different viewing angles, the range of viewing angles can be further enhanced by taking into account the desired coverage of angles in the training phase of the lying mirror, similar to our multi-wavelength lying mirror designs outlined earlier.

Looking forward, diffractive visual processor-based lying mirrors offer valuable potential across a wide range of fields and have diverse applications, from entertainment and experimental psychology to defense and security. They can also be used for disorienting adversaries by altering the perception of space and creating visual ambiguity. With the emerging advancements in nanofabrication techniques, this framework can be fabricated on various substrates to operate passively under spatially and temporally incoherent radiation. This scalability encourages further exploration into integrating optical computing hardware for more efficient and versatile visual data processing applications, expanding their utility across various fields, including anti-surveillance technologies.

## Methods

### Forward Model of a Lying Mirror

The forward model of a lying mirror employs coherent monochrome illumination at a wavelength of $\lambda$. The optical forward model of the diffractive processor can be generally described by two sequential processes: (1) free-space propagation of the light wave between two consecutive planes, and (2) modulation of the light wave by the diffractive layer or the structured mirror. The free-space propagation was modeled using the angular spectrum approach[30], expressed as:

$$u(x, y, z + d) = \mathcal{F}^{-1}\left\{\mathcal{F}\{u(x, y, z)\} \cdot H(f_x, f_y; d)\right\} \quad (2),$$

where $u(x, y, z)$ represents the original complex-valued field at the coordinate of $z$ along the optical axis, and $u(x, y, z + d)$ is the resulting light field at the coordinate of $z + d$ after propagating over an axial distance of $d$. $f_x$ and $f_y$ denote the spatial frequencies along the $x$ and $y$ directions, respectively. $\mathcal{F}$ and $\mathcal{F}^{-1}$ are the 2D Fourier transform and 2D inverse Fourier transform, respectively. $H(f_x, f_y; d)$ describes the free-space transfer function, defined as:

$$H(f_x, f_y; d) = \begin{cases} \exp\left\{jkd\sqrt{1 - \left(\frac{2\pi f_x}{k}\right)^2 - \left(\frac{2\pi f_y}{k}\right)^2}\right\}, & f_x^2 + f_y^2 < \frac{1}{\lambda^2} \\ 0, & f_x^2 + f_y^2 \geq \frac{1}{\lambda^2} \end{cases} \quad (3),$$

where $j = \sqrt{-1}$ and $k = \frac{2\pi}{\lambda}$.

The transmission coefficient of a diffractive feature at the lateral position $(x, y)$ was described by:

$$t(x, y) = \exp\{j\phi(x, y)\} \quad (4),$$

where $\phi(x, y)$ represents the phase modulation value for the corresponding diffractive feature, which forms the trainable parameters of the lying mirror.

The standard reflective mirror was assumed to be lossless with 100% reflectivity. In the structured mirror design shown in Fig. 1b, we applied the same modeling principles, treating it as a pure phase modulator, with a reflection coefficient of:

$$r(x, y) = \exp\{j\phi(x, y)\} \quad (5).$$

For the multi-wavelength experimental results and the corresponding lying mirror design, three illumination wavelengths ($\lambda = 480,\ 550$ and $600\ nm$) were utilized, with each following the corresponding forward model described above.

For the broadband experimental results and the corresponding lying mirror design, multiple illumination wavelengths were randomly sampled from $520\ nm$ to $570\ nm$ in each batch. Each wavelength followed the corresponding forward model described above.

With the above numerical forward model, we employed a deep learning-based approach to optimize the phase values of the diffractive layers. The training process involved feeding the input images through the optical system, calculating the output, and comparing it with the predefined target "dummy" output image to compute the loss. Then, by using automatic differentiation, we backpropagated the gradients to the trainable parameters (i.e., the phase modulation values of the diffractive layer). Leveraging a gradient descent-based optimization method, commonly used in training deep neural networks, we iteratively optimized the lying mirror design to ensure that the output images closely matched the predefined target images.

**Training Loss Function and Evaluation Metrics**

In our lying mirror framework, the diffractive processor converts the input data/information into a "dummy" image representation at its output plane. This operation T can be expressed as:

$$\mathbf{O} = \mathrm{T}(\mathbf{I}) \quad (6)$$

where $I$ denotes the input information and $O$ denotes the output image intensity of the lying mirror. Throughout the training phase, the lying mirror systems were trained to minimize the loss function $\mathrm{Loss_T}(\mathbf{O}, \mathbf{I_{target}})$, defined as:

$$\mathrm{Loss_T}(\mathbf{O}, \mathbf{I_{target}}) = 1 - \mathrm{PCC}(\mathbf{O}, \mathbf{I_{target}}) \quad (7).$$

This loss function was tailored to enhance the structural similarity between the lying mirror output $O$ and the predefined target image $\mathbf{I_{target}}$. The PCC score was calculated using:

$$PCC(\mathbf{O}, \mathbf{I_{target}}) = \frac{\sum\left(O(x,y) - \overline{O}\right) \cdot \left(I_{target}(x,y) - \overline{I_{target}}\right)}{\sqrt{\sum\left(O(x,y) - \overline{O}\right)^2 \cdot \left(I_{target}(x,y) - \overline{I_{target}}\right)^2}} \quad (8),$$

where $\overline{O}$ and $\overline{I_{target}}$ are the average intensity values of the output $\mathbf{O}$ and the target image $\mathbf{I_{target}}$, respectively.

For the multi-wavelength experimental demonstration of the lying mirror framework, the loss function was defined as:

$$Loss = \sum_{i=1}^{3} 1 - \mathrm{PCC}(\mathbf{O}_i, \mathbf{I_{target}}) \quad (9),$$

where $\mathbf{O}_i$ represents the output under the illumination wavelength $\lambda_i$ ($i.e.$, $480, 550, 600$ $nm$).

We also used the PSNR metric to assess the output image quality, which was defined as:

$$PSNR = 10\log_{10}\left(\frac{MAX^2_{\mathbf{I_{target}}}}{\mathrm{MSE}(\mathbf{O}, \mathbf{I_{target}})}\right) \quad (10),$$

where $MAX^2_{\mathbf{I_{target}}}$ is the maximum possible pixel value of the target "dummy" image (equal to 1), and $\mathrm{MSE}(\mathbf{O}, \mathbf{I_{target}})$ is the Mean Square Error between the output and target images.

**Simulation of Spatially Partially Coherent Illumination**

To model the propagation of a partially coherent field, we define the phase profile $\phi_i(x,y)$ of the incident field as[31]:

$$\phi_i(x,y) = \mathrm{mod}\left[\frac{2\pi}{\lambda} W(x,y) * G(x,y), 2\pi\right] \quad (11),$$

where '*' denotes the 2D convolution operation. Here, $W(x,y)$ follows a normal distribution and $G(x,y)$ represents a Gaussian smoothing kernel. We numerically tailored these phase profiles $\phi_i(x,y)$ to achieve the desired correlation lengths $C_\phi$, which were determined using the autocorrelation function $R_\phi(x,y)$:

$$R_\phi(x,y) = \mathcal{F}^{-1}\{|\mathcal{F}\{\varphi_i(x,y)\}|^2\} = \exp\left(-\frac{\pi(x^2+y^2)}{C_\phi}\right) \quad (12).$$

With Eq. (12), we numerically estimated $C_\phi$ based on 4096 randomly generated phase profiles $\phi_i(x,y)$, given specific $W(x,y)$ and $G(x,y)$.

The time-average intensity $\hat{O}$ of the lying mirror under spatially partially coherent illumination was computed as[27–29,32]:

$$\hat{O} = \langle \widehat{O_i} \rangle = \lim_{N_\phi \to \infty} \frac{1}{N_\phi} \sum_{i=1}^{N_\phi} \widehat{\mathbf{O}_i} \quad (13).$$

During the training of partially coherent lying mirror designs, we used $N_\phi^{train} = 16$ random phase profiles per input image, while the final performance evaluations employed $N_\phi^{test} = 4096$ random phase profiles for each test object/image.

**Experimental demonstration**

The structured mirror design was experimentally validated in the visible part of the spectrum. A fiber laser (Fianium ultrafast fiber lasers) was used as the illumination source, capable of emitting at $480, 550$ and $600\ nm$, either sequentially or simultaneously. The light was first filtered through a 4f system with a pinhole at the Fourier plane to improve the beam quality. It was then polarized using a linear polarizer to align with the operating direction of the SLM's liquid crystal array for object projection. The polarized light was modulated by an SLM (HOLOEYE LC-R 2500, pixel pitch: $19\ \mu m$, resolution: $1280 \times 960$) which displayed the input phase images. A phased light modulator (PLM, DLP6750Q1EVM, pixel pitch: $10.8\ \mu m$, resolution: $1358 \times 800$) was used as the optimized structured lying mirror to convert the input images into ordinary-looking "dummy" images. The output intensity was captured by a camera (QImaging Retiga-2000R, pixel pitch: $7.4\ \mu m$, resolution:$1600 \times 1200$). The distance between the input object plane (SLM) to the structured mirror plane was $136.8\ mm$, and the distance between the structured mirror to the sensor plane was $145.0\ mm$.

For the broadband lying mirror experiments, the same illumination source was used as before; we also used two SLMs for the object plane and the structured lying mirror plane (Meadowlark XY Phase Series, pixel pitch: $8 \mu m$, resolution: $1920 \times 1200$; HOLOEYE PLUTO-2.1, pixel pitch $8 \mu m$, resolution $1920 \times 1080$). The axial distance between the input object plane (SLM) to the structured mirror plane was $120.1\ mm$, and the distance between the structured mirror to the sensor plane was $96.4\ mm$.

**Supplementary Information** includes:

- **Parameters and Digital Implementation for Numerical Analyses**
- **Supplementary Figures S1-S11**

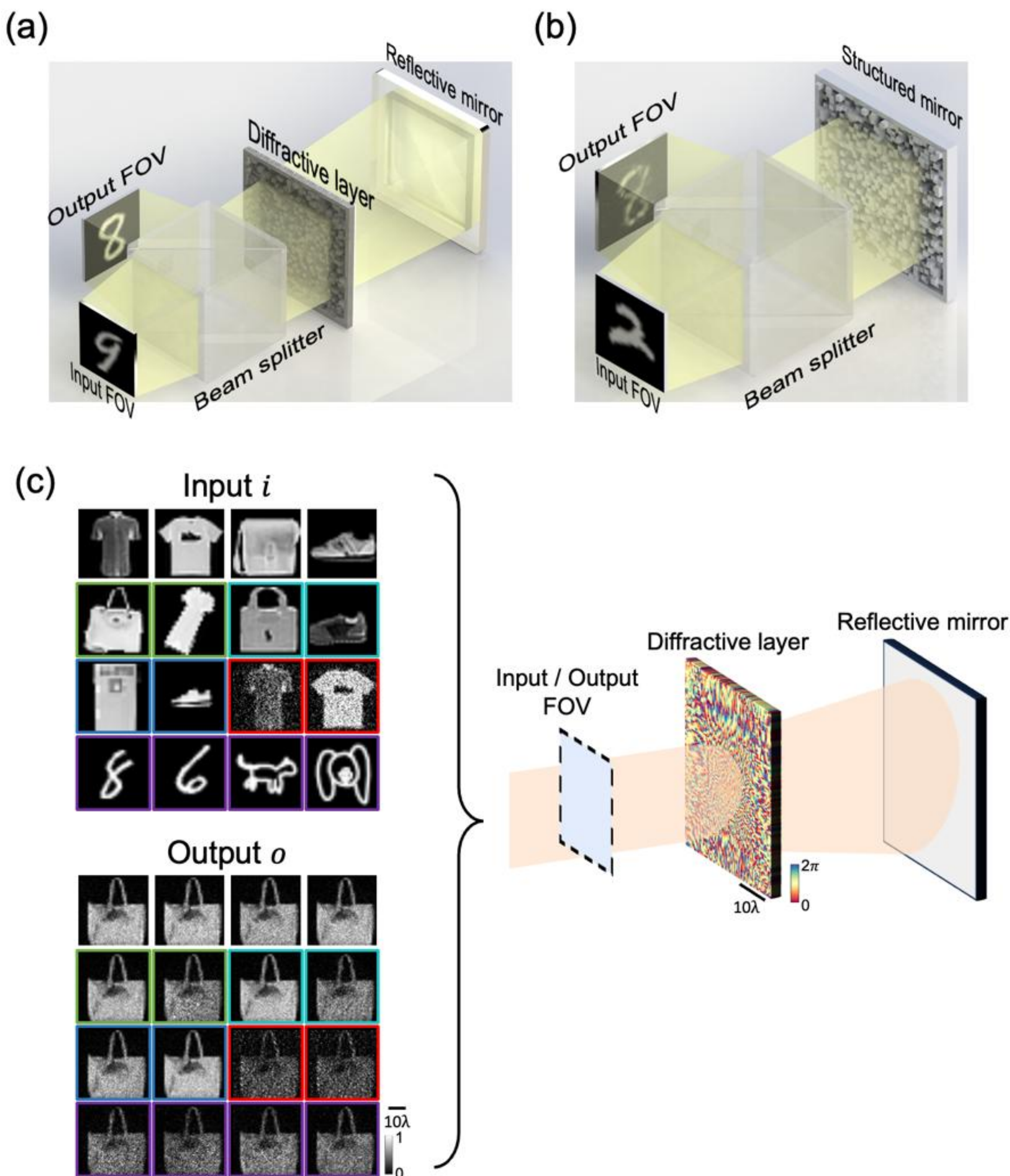

**Figure 1. All-optical lying mirror**. (a) Schematic of the lying mirror, consisting of a diffractive layer paired with a standard reflective mirror. (b) Schematic of a structured mirror designed for optical image concealment. (c) The lying mirror effectively conceals arbitrarily selected input images, transforming them into ordinary-looking “dummy” patterns that mislead human observers under various conditions, including random rotation, scaling, shifting, noise, and external images beyond the training dataset. Color coding in (c) refers to randomly rotated (green), scaled (blue), shifted (cyan) and noise-added (red) images as well as external image distributions (purple).

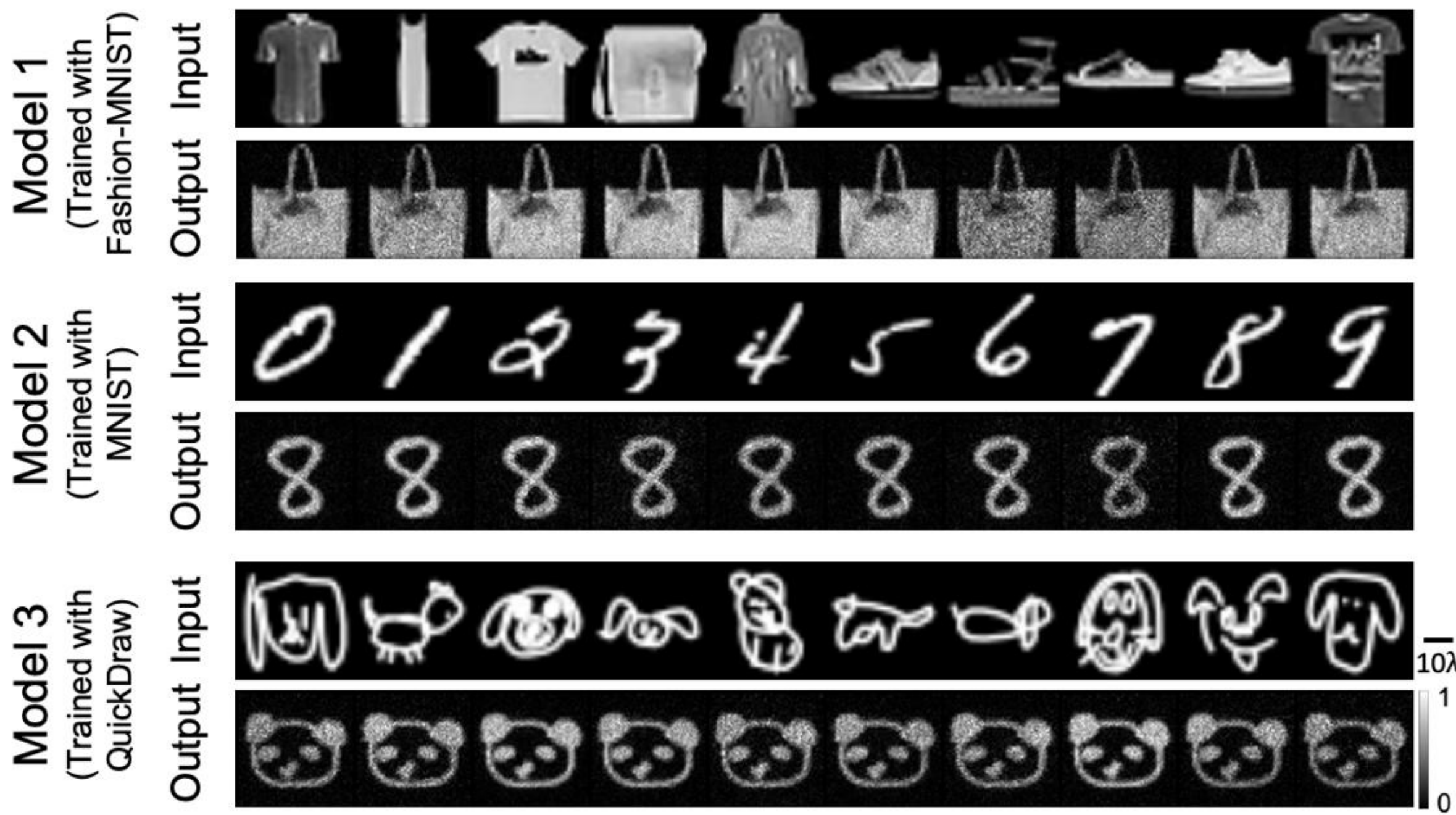

**Figure 2. Blind testing results of lying mirrors.** Three lying mirror models, trained with Fashion-MNIST, MNIST, and QuickDraw datasets, respectively, demonstrate the capability of the lying mirror framework. The system transforms various randomly selected input images—such as clothing items, handwritten digits, or animals—into ordinary-looking images resembling a "bag", a handwritten digit "8", or a "panda", respectively.

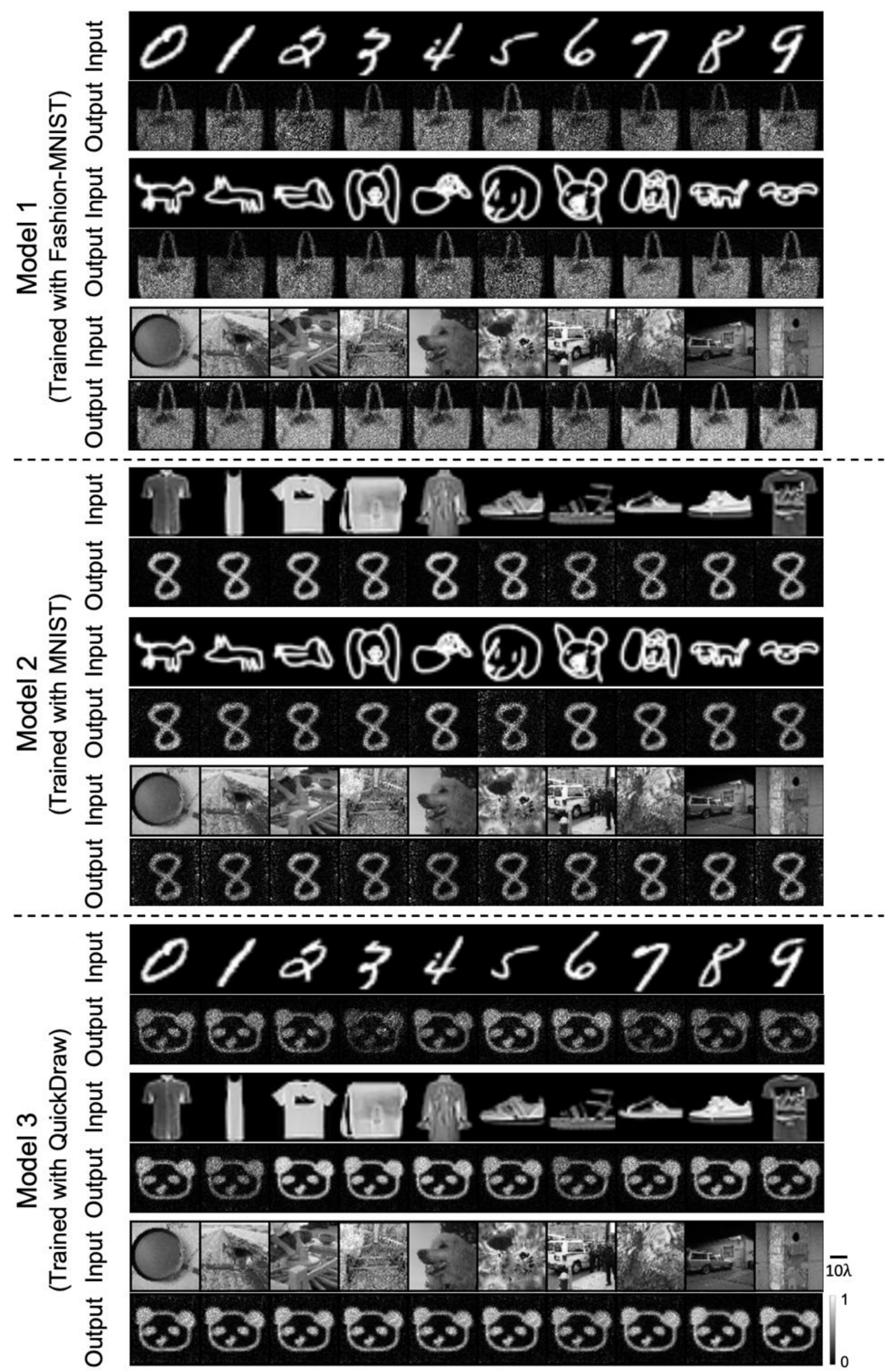

**Figure 3. Blind testing of the external generalization ability of lying mirrors.** Lying mirrors, each trained on one image dataset, were randomly tested on three additional image datasets, including the ImageNet dataset, that were not involved in the training process, to evaluate their ability to generalize across different data types never seen before.

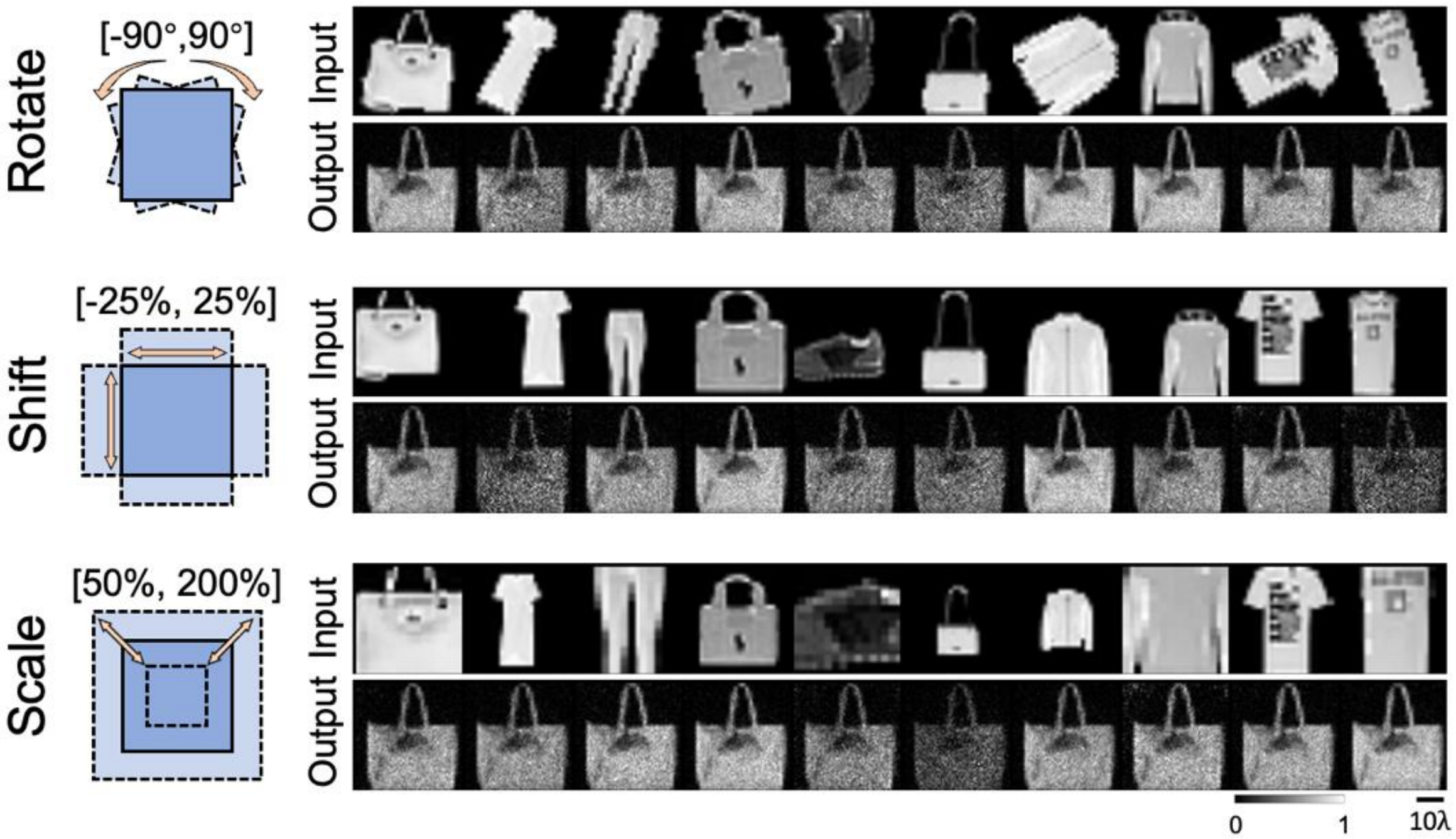

**Figure 4. Evaluation of the performance of a lying mirror under various attacks.** The lying mirror showed robustness against different types of input data manipulations, including random rotations ranging from -90° to 90°, random shifts ranging from -25% to 25% of the input image size, and random scaling from 50% to 200% of the original image size.

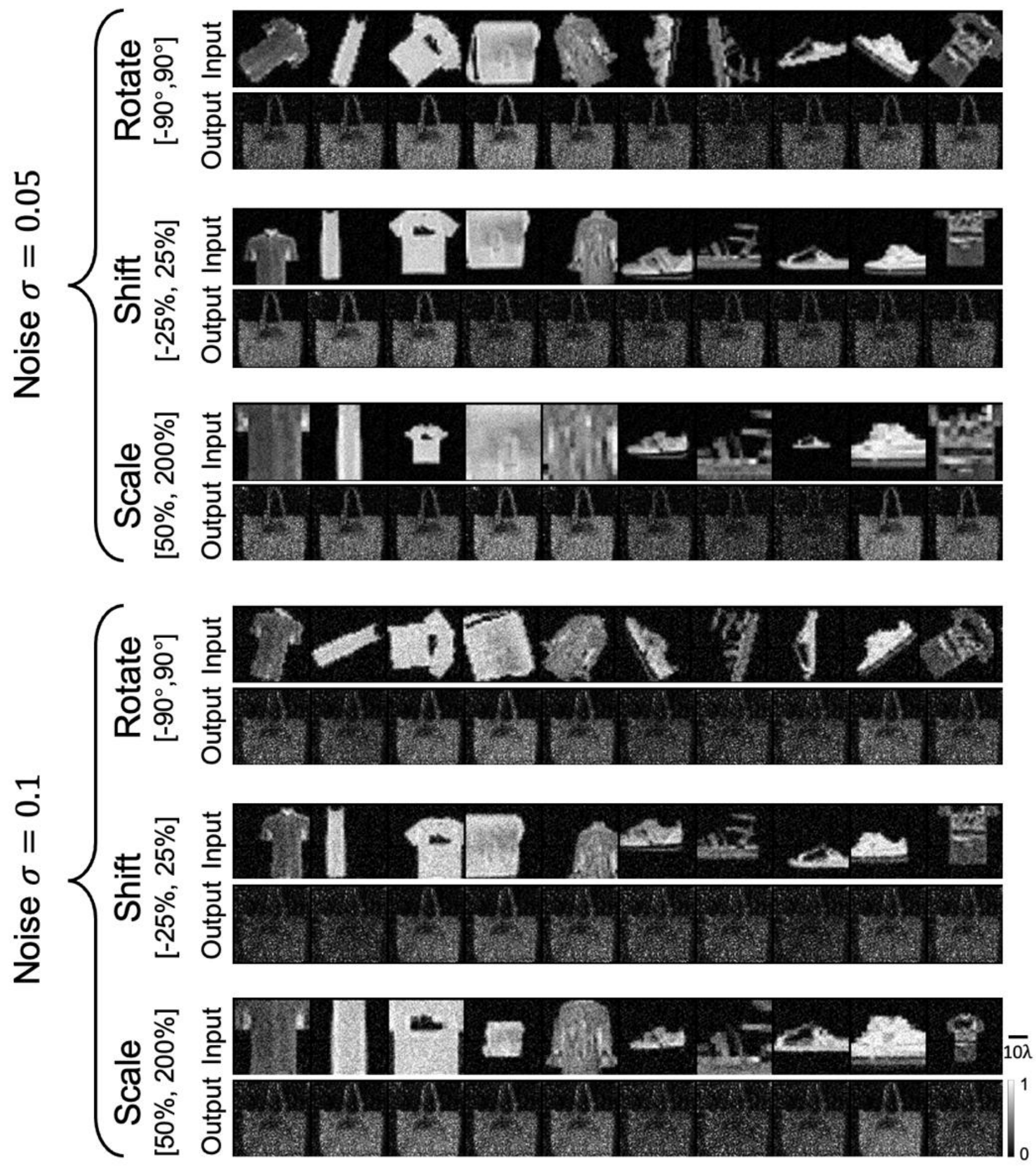

**Figure 5. Evaluation of the performance of a lying mirror under noise.** The lying mirror was tested with images subjected to different noise levels ($\sigma = 0.05$ and $\sigma = 0.1$) as well as random image rotations, shifts and scaling with the same amounts as reported in Fig. 4.

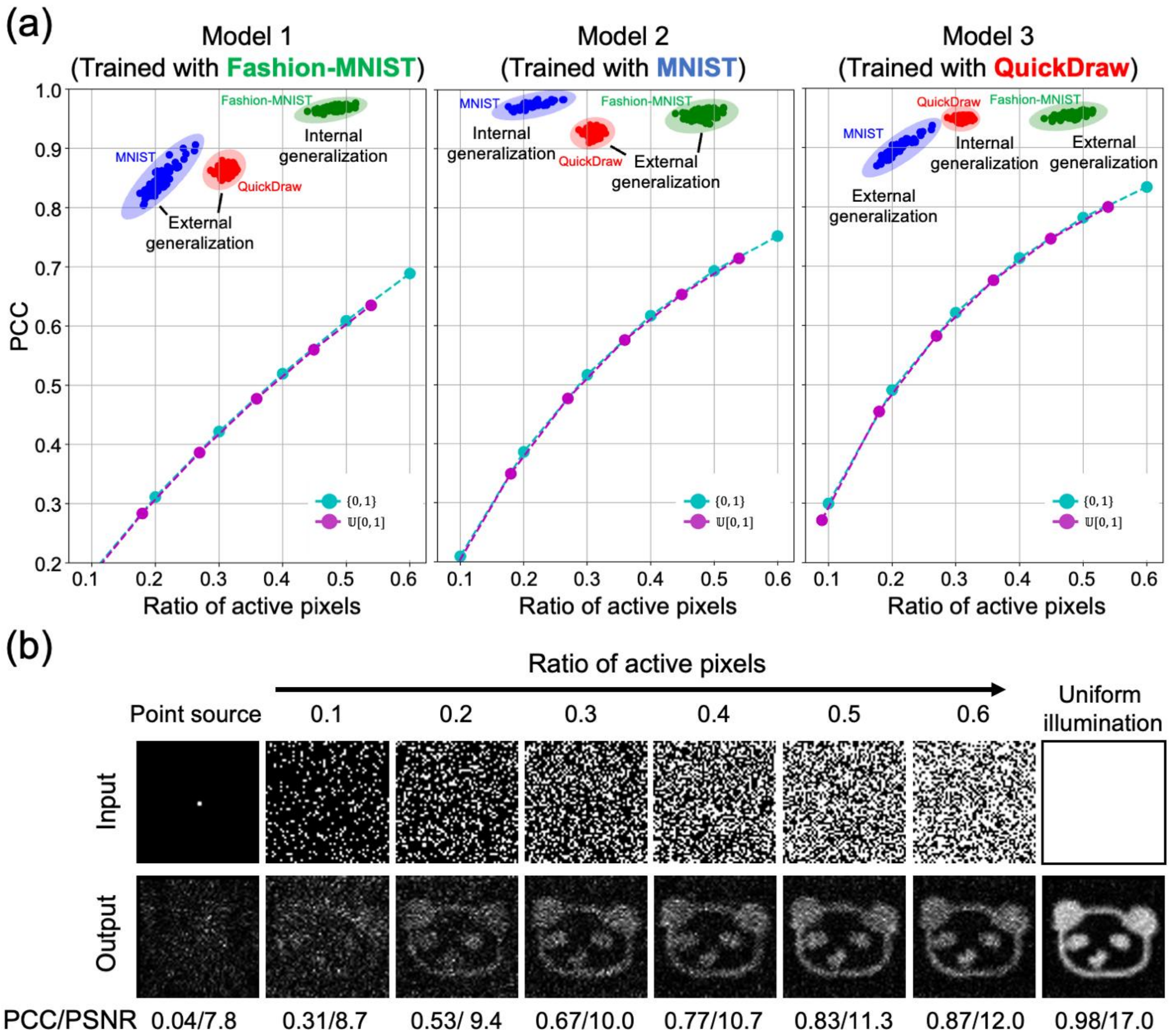

**Figure 6. Quantitative analysis of the performance of lying mirror outputs for different input images.** (a) Three lying mirror models, trained with the Fashion-MNIST, MNIST, and QuickDraw datasets, were blindly evaluated using these 3 image datasets as well as binary random patterns and random patterns with uniform distribution $\mathbb{U}[0,1]$. (b) Visualization of Model 3 output images with different ratios of active pixels. The results using a point source and uniform illumination are also included.

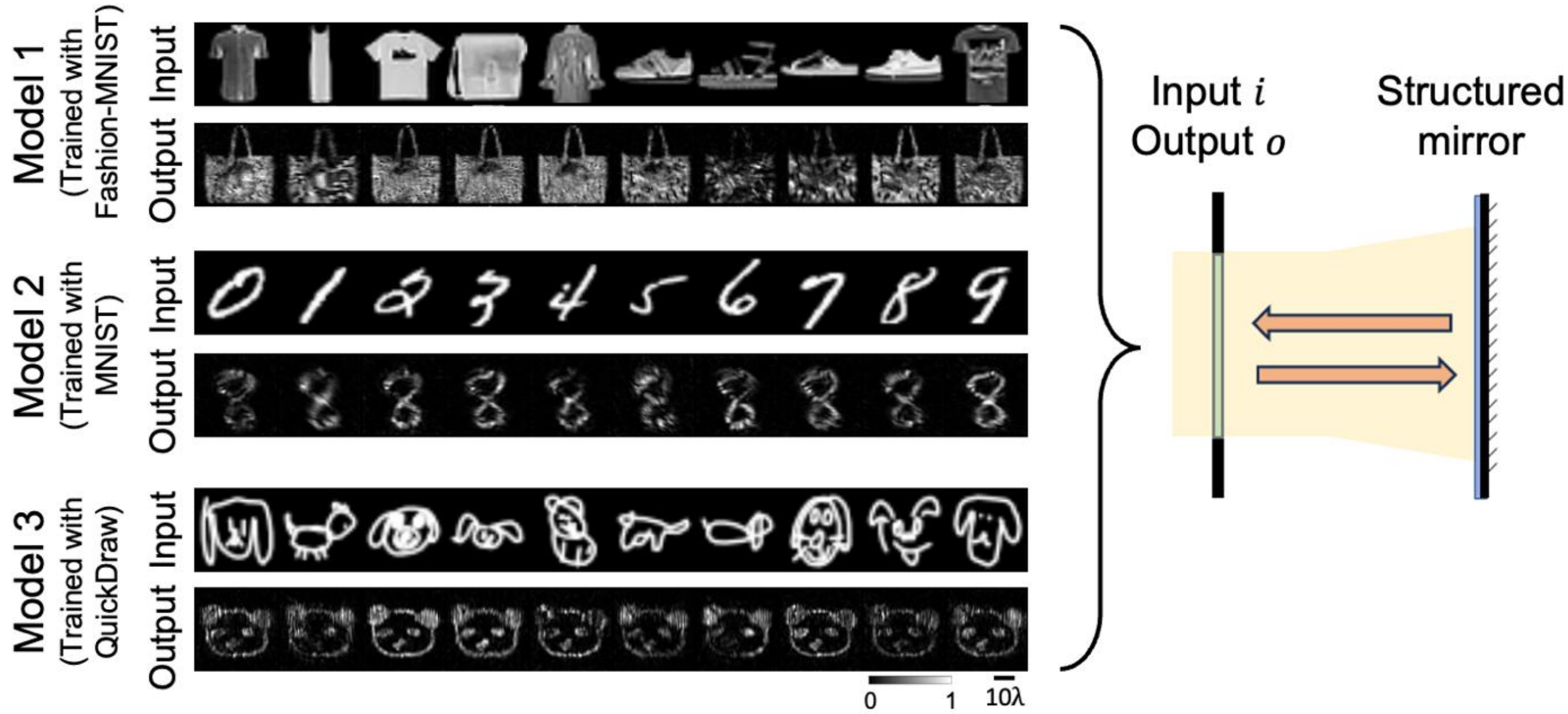

**Figure 7. Schematic of a structured lying mirror, a variant of the lying mirror framework.** This setup comprises a structured lying mirror with phase modulation, integrating the diffractive layer directly with a standard reflective mirror. Testing results of three models, trained with Fashion-MNIST, MNIST, and QuickDraw datasets, show that the structured lying mirror can transform a variety of randomly selected input images—clothing items, handwritten digits, or animals—into ordinary-looking “dummy images resembling a “bag”, a handwritten digit “8”, or a “panda”, respectively.

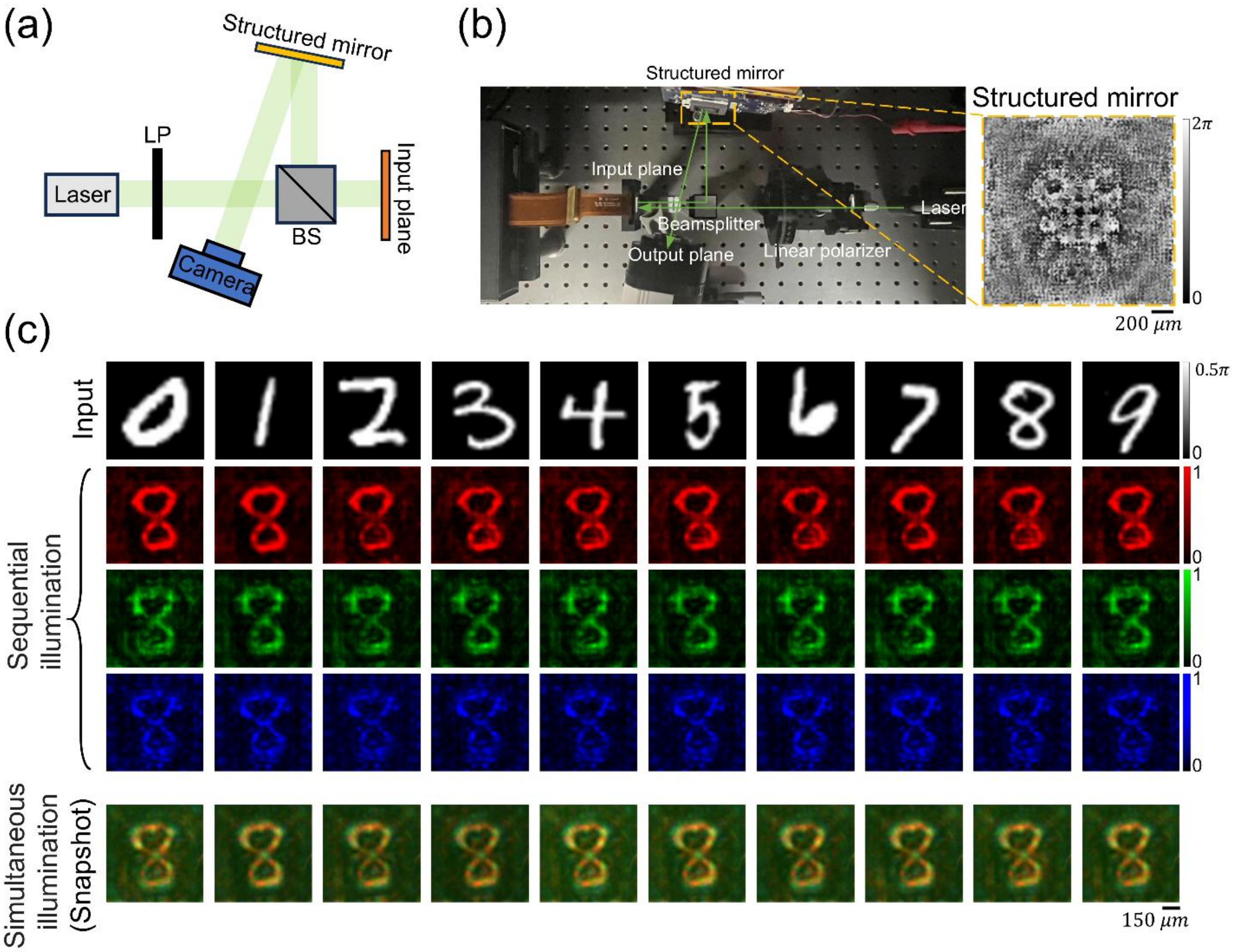

**Figure 8. Experimental demonstration of a structured lying mirror at the visible part of the spectrum.** (a) Schematic of the multi-wavelength lying mirror experimental setup with $\lambda = 480, 550$, and $600\ nm$ illumination. BS: Beamsplitter; LP: Linear polarizer. (b) The photograph of the experimental setup and the optimized phase pattern of the structured mirror, implemented using a MEMS-based micro-mirror array. (c) Experimental results of the structured lying mirror, converting phase-only randomly selected objects into the target "dummy" image (handwritten digit '8') for three illumination wavelengths. The structured mirror was illuminated by these wavelengths, either sequentially or simultaneously (bottom row).

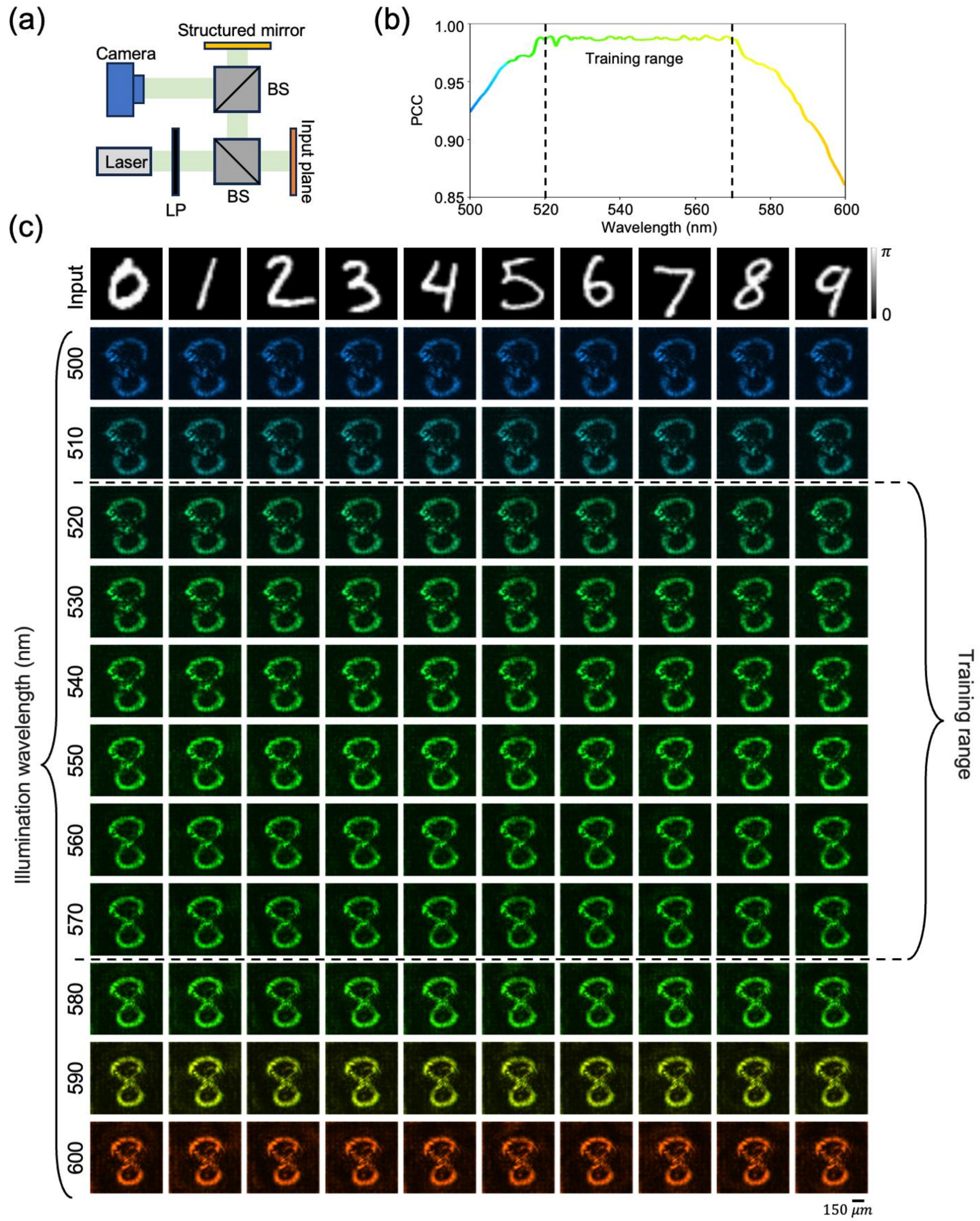

**Figure 9. Experimental demonstration of a structured lying mirror for broadband operation.** (a) Schematic of the experimental setup for the broadband lying mirror. (b) Numerical performance analysis

of the lying mirror across the spectral range of $500-600\ nm$. (c) Experimental results of the lying mirror under $\{500, 510, \ldots, 600\}\ nm$ illuminations.